# Early Diagnosis of Alzheimer's Diseases and Dementia from MRI Images Using an Ensemble Deep Learning

Mozhgan Naderi[1], Maryam Rastgarpour*[2] , Amir Reza Takhsha[3]

**Abstract**

Alzheimer's Disease (AD) is a progressive neurological disorder that can result in significant cognitive impairment and dementia. Accurate and timely diagnosis is essential for effective treatment and management of this disease. In this study, we proposed two low-parameter Convolutional Neural Networks (CNNs), IR-BRAINNET and Modified-DEMNET, designed to detect the early stages of AD accurately. We also introduced an ensemble model that averages their outputs to reduce variance across the CNNs and enhance AD detection. Both CNNs are trained, and all models are evaluated using a Magnetic Resonance Imaging (MRI) dataset from the Kaggle database. The dataset includes images of four stages of dementia, with an uneven class distribution. To mitigate challenges stemming from the inherent imbalance in the dataset, we employed the Synthetic Minority Over-sampling Technique (SMOTE) to generate additional instances for minority classes. In the NO-SMOTE scenario, despite the imbalanced distribution, the ensemble model achieved 98.28% accuracy, outperforming IR-BRAINNET (97.26%) and Modified-DEMNET (95.54%), with Wilcoxon p-values of 2.9e−3 and 5.20e-06, respectively, indicating significant improvement in correct predictions through the use of the average function. In the SMOTE scenario, the ensemble model achieved 99.92% accuracy (1.64% improvement over NO-SMOTE), IR-BRAINNET reached 99.80% (2.54% improvement), and Modified-DEMNET attained 99.72% (4.18% improvement). Based on the experimental findings, averaging the models' outputs enhanced AD diagnosis in both scenarios, while the diversity in the dataset introduced by SMOTE-generated instances significantly improved performance. Furthermore, the compact models we proposed outperformed those from previous studies, even in the presence of an imbalanced distribution.

## 1 Introduction

Alzheimer's Disease (AD) the most prevalent type of dementia, is distinguished by a worsening of behavioral symptoms, a decline in short-term memory, and a gradual deterioration in cognitive function [1]. Dementia can arise from various conditions and injuries that affect the brain. Among these, Alzheimer's disease is the most common form, accounting for approximately 60-70% of cases. [2]. Globally, the rising prevalence of AD and its associated challenges have garnered significant attention due to its increasing impact on society. As of 2023, the global count of individuals affected by dementia has surpassed 50 million [3], a substantial increase from 20.3 million in 1990[4]. In 2050, it is estimated that there will be 13.8 million individuals suffering from AD dementia, with 7.0 million of them aged 85 years or older [5].

Advanced imaging methods, including Magnetic Resonance Imaging (MRI) as well as Positron Emission Tomography (PET), scans with tracers such as Fluoro-Feoxy-D-Glucose (FDG) and Pittsburgh Compound B (PiB), reveal distinctive brain changes in patients with AD, not only in advanced stages but also in early and even pre-symptomatic phases, which aid in the diagnosis of AD's underlying pathological processes [8]. MRI techniques uncover specific patterns of brain damage that distinguish AD from other neurological conditions. Additionally, they identify brain abnormalities associated with the transition from MCI to AD and other behavioral outcomes [9].

Detecting AD in its early stages is critical for developing effective treatments and interventions, as the disease shows greater responsiveness to treatment when identified early on[10]. Extensively extracting features from MRI images or medical records to diagnose brain soft tissues for early dementia and AD prediction demands significant time and effort, with the potential for errors escalating due to the similarity between soft and healthy tissues in MRI scans [11]. Considering this complexity, the Deep Learning framework, stemming from machine learning, has been investigated as a feasible solution. This is because of the innate capacity of Deep Learning models to efficiently process complex input data without relying on time-consuming and poorly scalable feature extraction procedures [12].

---
[1] Faculty of Engineering, Department of Computer Engineering, Islamic Azad University, Saveh, Iran
  *mozhganaderii@gmail.com, m.naderi@stu.iau-saveh.ac.ir*
[2] Faculty of Engineering, Department of Computer Engineering, Islamic Azad University, Saveh, Iran
  *m.rastgarpour@iau-saveh.ac.ir*; *m.rastgarpour@gmail.com*
[3] Faculty of Engineering, Department of Computer Engineering, Islamic Azad University, Saveh, Iran
  *amirrezatakhsha@gmail.com, amirreza.takhsha@iau.ac.ir*

Deep learning has recently become the leading approach in medical imaging, particularly in the detection of Alzheimer's disease [13]. Recent advancements in Deep Learning have enabled the identification of distinct patterns in the progression of AD [14]. Deep neural networks excel at identifying subtle and complex changes in brain structure from data, enabling the monitoring of AD progression and contributing to reliable diagnostic outcomes [15]. The application of Deep learning improves efficiency and speed in the detection process, which are crucial for the early diagnosis of patients and timely medical treatment [16].

Convolutional Neural Networks (CNNs), a subset of deep learning models, excel at extracting features efficiently, making them indispensable in domains like medical imaging. Designed for image analysis, CNNs stack layers such as convolutional, pooling, activation, and fully connected layers—to capture complex features while minimizing parameters compared to traditional neural networks [17]. They leverage techniques like local receptive fields, shared weights, filters, strides, and padding to extract features, reduce dimensionality, introduce non-linearity, and predict label probabilities [17]. By processing 2D or 3D images, CNNs automatically learn meaningful local and global features, avoiding errors common in manually crafted features [18].

In this study, we leveraged deep learning to accurately detect AD stages by introducing two CNN-based models: IR-BRAINNET and Modified-DEMNET. These models utilize convolutional operations on MRI images for automatic feature extraction to classify AD stages.

The parameters of a CNN model significantly impact its performance [19]. While models with more parameters may achieve better results, they also increase computational complexity and memory requirements, posing challenges for practical implementation [20]. Since many hospitals lack robust computational resources [21], reducing computational demands is vital for effective medical image analysis in resource-constrained settings. Compact networks ensure efficient processing with limited resources and integrate seamlessly into portable edge devices [22], offering flexibility, reduced data exchange, and lower memory usage. For instance, compact CNNs can be deployed on Android devices to detect AD stages, demonstrating their practical real-world applications.

In our approach, we designed two CNNs with a low number of parameters to ensure effective detection of AD stages. To enhance the performance of IR-BRAINNET, we utilized transfer learning, leveraging pre-trained knowledge from another domain. This technique, which employs both frozen weights (unchanged during training) and fine-tuned weights (adjusted for specific tasks), outperforms randomly initialized filters, even when sourced from unrelated tasks, thereby improving overall performance [25]. Specifically, we transferred convolutional layer weights from the pre-trained VGG-19 model, trained on the ImageNet dataset [26], to a convolutional layer in IR-BRAINNET.

To further enhance AD stage detection, we employed ensemble learning techniques to leverage the strengths of both proposed CNN models. Ensemble learning, a widely used approach, combines diverse CNN architectures to achieve better predictive performance than individual models [23]. By integrating multiple models, ensemble learning improves the robustness and overall accuracy of learning systems [24]. In deep learning, where architectures often exhibit high variance and low bias, simple averaging of predictions across ensemble models effectively reduces variance, improving generalization performance [24]. Accordingly, we combined predictions from both CNN models using an averaging function, forming a robust ensemble model.

We used the Kaggle Alzheimer's dataset [27] to train and evaluate the two CNN models and to assess the performance of our proposed ensemble model. This dataset categorizes samples into four classes based on AD severity: Moderate Demented (MOD), Mild Demented (MID), Very Mild Demented (VMD), and Non-Demented (ND). However, a significant challenge with this dataset is its pronounced class imbalance, which can cause models to over-predict the majority class, leading to frequent misclassification of minority classes [28]. Addressing this imbalance is essential for ensuring accurate and fair model predictions.

To tackle issues arising from the imbalanced dataset, which impacts the performance of the models, we implemented the Synthetic Minority Over-sampling Technique (SMOTE). SMOTE, as a highly effective approach, generates additional instances from the minority class to bring the dataset size in line with that of the majority class [29]. SMOTE is the preferred method for handling imbalanced datasets and it is both effective and easy to implement [30]. Our aim with the utilization of SMOTE was to align the dataset sizes of other classes with that of the majority class, ND, and enhance the models' effectiveness in learning patterns of minority classes, especially MOD. This strategy can be highly effective in medical imaging classification, especially when accurately categorizing different abnormality stages is crucial given their imbalanced distribution, to achieve accurate diagnosis and optimal treatment.

The major contributions of this work are as follows:

- In this paper, we proposed two low-parameter CNN models: IR-BRAINNET, developed from scratch, and a more compact version of the DEMNET CNN [7]. Designed for low memory consumption and operational efficiency, these models aim to accurately detect the four stages of AD using the Kaggle Alzheimer's dataset, which features a highly imbalanced distribution.

- This paper employed a transfer learning technique by transferring weights from a convolutional layer of the pre-trained VGG-19 model to a convolutional layer in IR-BRAINNET, mitigating the impact of limited training data and enhancing model performance by leveraging learned features from an early layer of the VGG-19 model.

- We utilized an ensemble learning strategy by employing the average function, which is an efficient computational function, to combine the prediction vectors from IR-BRAINNET and Modified-DEMNET. By reducing variance across models and taking advantage of the strengths of both CNNs, this approach can enhance AD detection performance.

- To address the class imbalance in the Kaggle Alzheimer's dataset and improve AD diagnosis, we applied the SMOTE function to generate additional instances for the minority classes. Incorporating interpolated synthetic data for the minority classes improves the models' ability to learn the underlying patterns of these classes and mitigates the risk of overfitting.

## 2 Related Works

AD, a neurodegenerative disorder impacting millions globally, highlights the critical need for early detection and prognosis to enable effective disease management. Recently, Deep Learning methods have gained prominence for early AD identification, utilizing various neuroimaging modalities such as MRI scans. In this section, we present an overview of studies from the literature and efforts in this field, as we propose models for detecting stages of Alzheimer's disease within the Deep Learning framework.

Murugan et al. [7] developed DEMNET, a CNN model for classifying categories in the Kaggle Alzheimer's dataset. To address data imbalance, they applied SMOTE. DEMNET includes four blocks, each with two convolutional layers, batch normalization, and max pooling, followed by dense layers with 512, 128, and 64 neurons. Softmax activation in the final layer produces class probabilities, while dropout layers counter overfitting. DEMNET achieved 95.23% accuracy and 96% precision with SMOTE, compared to 85% accuracy and 80% precision without it.

Sadiq Fareed et al. [31] introduced ADD-NET, a CNN model for identifying Alzheimer's disease from MRI images in the Kaggle Alzheimer's dataset. ADD-NET consists of four convolutional blocks, each with a convolutional layer, ReLU activation, and average pooling, with filter counts doubling from 16 to 128. After the convolutional layers, ADD-NET has a flatten layer, a dense layer with 256 neurons, a dropout layer, and a final dense layer with Softmax for probability output. To address data imbalance, they used SMOTETOMEK, a technique combining SMOTE and Tomek links, achieving 97.05% accuracy and 97% precision. Without SMOTETOMEK, the model's precision for the MOD class dropped to 0%.

Fathi et al. [32] proposed an ensemble Deep Learning model for the early diagnosis of AD utilizing MRI images from the ADNI dataset. Their methodology involved leveraging six established CNNs: DenseNet-201, DenseNet-169, DenseNet-121, ResNet50, Inception-ResNet v2, and VGG-19. Each CNN underwent individual training and testing on the dataset. Following this, they employed a weighted probability method to amalgamate the probabilities predicted by each classifier. The weight assigned to each classifier in the final ensemble model was determined based on its accuracy after the training phase. The proposed ensemble model achieved accuracy rates across various classification groups as follows: 98.57% for NC/AD, 96.37% for NC/EMCI, 94.22% for EMCI/LMCI, 99.83% for LMCI/AD, 93.88% for NC/EMCI/LMCI/AD, and 93.92% for NC/MCI/AD.

Mehmood et al. [33] used a transfer learning approach with a modified VGG-19 model to detect Mild Cognitive Impairment (MCI) from MRI images in dementia research. Focusing on gray matter segmentation, they aimed to classify Normal Control (NC), Early MCI (EMCI), Late MCI (LMCI), and Alzheimer's Disease (AD) in the ADNI dataset. They divided VGG-19 into two configurations: in Group 1, the first three blocks were frozen; in Group 2, the first four blocks were frozen. After customizing the final layers, the model achieved 93.83% accuracy without data augmentation 95.38% with it in Group 1, 95.33% without augmentation, and 98.73% with it in Group 2, for AD vs. NC.

Helaly et al. [34] used both 2D and 3D brain scans from the ADNI dataset to classify Alzheimer's disease (AD) stages. They applied simple CNN models with three convolutional layers (32 filters in the first, 64 in the next two), three max-pooling layers, two dropout layers, a flatten layer, and two fully connected layers with ReLU activation. Softmax activation in the final layer classified four AD stages: AD, EMCI, LMCI, and NC. For 3D images, 3D convolutional layers with a 3×3×3 kernel were used, while 2D images had 3×3 kernels. The models achieved 93.60% and 95.17% accuracy for 2D and 3D scans, respectively. Fine-tuning a pre-trained VGG-19 further improved accuracy to 97% for multi-class classification.

Nawaz et al. [35] developed a CAD system for real-time Alzheimer's disease (AD) diagnosis, addressing class imbalance and overfitting through transfer learning. Using a pre-trained AlexNet model to extract deep features, they classified these features with SVM, KNN, and RF. The model achieved 99.21% accuracy with SVM, 57.32% with KNN, and 93.97% with RF, demonstrating high accuracy in distinguishing between the four classes (Normal, VMD, MID, MOD) in the Oasis dataset when using SVM. The AlexNet-based model outperformed both handcrafted features and a custom CNN, proving the effectiveness of deep features for AD stage classification.

Rashid et al. [36] introduced Biceph-net, a lightweight AD diagnosis model based on the VGG-16 architecture, using 2D MRI scans. Biceph-net includes a unique Biceph Module after the VGG-16 flatten layer, which splits into a triplet branch (applying triplet loss) and a concatenate branch (using cross-entropy loss). The module arranges 2D MRI slices from the same subject together while distinguishing them from other subjects. During testing, the triplet branch groups similar slices for effective classification. The model achieved 100% accuracy for Cognitively Normal (CN) vs. AD, 98.16% for Mild MCI vs. AD, and 97.80% for differentiating CN, MCI, and AD.

Arafa et al. [15] developed a CNN to classify MID and ND in the Kaggle Alzheimer's dataset. The model has three 2D convolutional layers with 3×3 filters, using 8, 16, and 32 filters respectively, each followed by pooling layers (one with 3×3 and two with 2×2 sizes). The classification layers include two dense layers with ReLU, a final dense layer with sigmoid activation, and dropout layers to prevent overfitting. They also fine-tuned a pre-trained VGG-16, freezing the first 16 layers and adjusting the last convolutional layers. The CNN achieved 99.98% accuracy with an 80/20 split and 99.9% with a 70/30 split, while the VGG-16 model reached 97.44% with a 60/40 split.

Roopa et al. [37] used a Teaching-Learning-based Brainstorm (TLBS) segmentation method to extract important features from MRI images for Alzheimer's classification. TLBS, which combines learning and associative search agents, aids in finding optimal solutions for feature extraction. These features were input into a deep CNN with two convolutional layers, two max-pooling layers, and a Softmax-activated fully connected layer. The CNN parameters were fine-tuned using TLBS, achieving 97.03% accuracy on the ADNI database (Normal, MCI, AD) and 95.80% on the Kaggle Alzheimer's dataset (four categories).

El-Assy et al. [38] proposed a concatenated ensemble model based on two different convolutional architectures for detecting instances from the ADNI dataset into five distinct classes: CN, EMCI, LMCI, AD, and MCI. The first CNN in this ensemble consists of five convolutional layers, using filters in the sequence of 16, 64, and 256, along with three max-pooling layers and a batch normalization layer. The second CNN follows the same structure but employs a higher number of filters for each layer: 64, 256, and 512. They concatenated the 128-dimensional feature vectors extracted from the fully connected layers of these architectures. The combined features were then fed into a prediction layer consisting of 5 neurons. The proposed ensemble model achieved an accuracy of 99.13% in five-class classification.

Sorour et al. [39] introduced three CNN models and a pre-trained VGG-19 model to classify Non-Demented and Moderately Demented stages in Kaggle's Alzheimer's dataset. The first CNN has three convolutional layers with 16, 32, and 64 filters, each followed by max pooling, and two dense layers with 128 and 64 neurons, ending with a Softmax layer for classification. They also proposed a CNN-LSTM model, combining CNN-based feature extraction with an LSTM layer containing 100 units. Additionally, they introduced a CNN-SVM model, which uses an SVM classifier on features extracted from a CNN. For VGG-19, they applied SVM on the feature map. The CNN-LSTM model achieved the highest accuracy, 99.92%, with data augmentation.

Odimayo et al. [40] developed a model called SNeurodCNN to identify instances of AD and MCI, including both progressive and stable forms, within the ADNI dataset. The SNeurodCNN architecture features three convolutional layers with filter sizes of 32, 32, and 64, respectively, and includes max pooling layers after the first and third convolutional layers. In its fully connected section, the network comprises two dense layers: the first with 500 neurons, and the final layer with 2 neurons for distinguishing between MCI and AD. The model demonstrated high performance, achieving an accuracy of 97.8% in differentiating AD from MCI using the para-sagittal viewpoint and 98.1% using images from the mid-sagittal viewpoint.

Sekaran C et al. [41], IC-Net, a segmentation model based on U-Net, was introduced for enhanced brain tumor segmentation in MRI images. It incorporates Multi-Attention (MA) blocks that highlight critical regions by sequentially applying convolution, batch normalization, and ReLU activation. Feature Concatenation Networks (FCNs) combine features from different levels using convolution layers with ReLU, Tanh, and Sigmoid activations, concatenated and processed through a final ReLU convolution. Additionally, an Attention block with parallel 1×1 convolutions filters out irrelevant details, generating an attention mask that re-weights features to sharpen focus. IC-Net achieved superior performance over U-Net with training and validation accuracies of 99.65% and 97.94% on the BraTS 2020 dataset.

Yuhuan Hu et al. [42] developed a method to differentiate AD patients from normal controls by combining resting-state fMRI (rs-fMRI) and structural MRI (sMRI) features. sMRI was used to analyze cerebral cortex properties, while rs-fMRI focused on graph metrics in functional networks. The Multivariate Minimal Redundancy Maximal Relevance (MRMR) algorithm selected optimal features, maximizing statistical relevance. The classification was performed with an SVM using leave-one-out cross-validation, achieving 78.72% accuracy for rs-fMRI, 87.23% for sMRI, and 91.49% when both modalities were combined.

Geneedy et al. [18] developed a CNN-based approach for AD identification using T1-weighted MRI scans from the OASIS dataset. Their low-parameter CNN model has five convolutional layers with max pooling, followed by a fully connected layer with dropout and a dense layer with Softmax activation, classifying images into four categories: VMD, MID, ND, and MOD. With 4,619,524 trainable parameters and data augmentation (brightness, zoom, rotation adjustments), the model achieved a 99.68% accuracy.

# 3 Material and Methods

In this section, we provide a concise overview of our proposed model. We then delve into the Kaggle Alzheimer's dataset, discussing its distribution and inherent imbalance. Subsequently, we elucidate our preprocessing steps, oversampling techniques, and the data splitting process. Following that, we explore the architectures and parameters of our two proposed CNN models: IR-BRAINNET and Modified-DEMNET, with the latter being a modified variant of the DEMNET model [7]. Finally, we elaborate on our proposed ensemble model, which combines the final Softmax outputs of the two CNN models, and provide insights into its functionality through the presentation of pseudocode for the proposed ensemble model.

## 3.1 Method Overview

Deep Learning has made significant strides in recent years, with numerous applications in various fields, such as healthcare, energy management, and image analysis. One such example is the utilization of CNNs for accurately identifying AD stages. As outlined in the related work section, CNNs have demonstrated great potential in automatically extracting complex features from MRI images of patients and effectively utilizing them for classifying AD stages. Within the realm of Deep Learning framework, we introduced two CNN models: IR-BRAINNET, a novel CNN model developed by us, and Modified-DEMNET, an adapted version of the DEMNET [7] model with reduced parameters, specifically tailored for identifying AD from MRI images. To augment the detection efficacy in our task, we harnessed the capabilities of two CNNs by designating them as individual models within a unified set. This unified set serves as the basis of the proposed ensemble model. The core ensembling operator of the proposed ensemble model utilizes an average function to fuse the information from individual elements of the ensemble set by averaging the likelihoods generated by the two proposed CNNs. For AD diagnosis, our study utilizes MRI images sourced from the Kaggle platform, similar to previous studies [7, 15, 27] that also employed the Kaggle Alzheimer's dataset as the main dataset for their research. As the Kaggle Alzheimer's dataset is inherently imbalanced, we employed the SMOTE technique to improve AD stage detection by generating synthetic images. This approach balances the dataset and can potentially mitigate model biases. However, due to the significant computational resources required by the SMOTE technique, we designed two CNN models to effectively identify AD stages while dealing with the imbalanced dataset. Additionally, employing an ensembling technique further enhances detection performance, allowing us to achieve better results in this challenging situation.

The objectives of this study include: (1) achieving optimal classification of Kaggle Alzheimer's dataset images, with a specific emphasis on effectively classifying the minority class MOD, along with MID, VMD, and ND instances, even without oversampling; (2) generating samples through synthetic minority oversampling to balance the dataset to improve AD stage detection; (3) introducing a novel CNN model with low-level parameters, coupled with modifications to an existing CNN model, DEMNET [7], aimed at early detection of AD stages with minimal resource usage while achieving remarkable performance; (4) applying an ensembling technique that combines the probability distribution vectors of the two proposed CNN models using a linear function to average all input vectors to enhance AD detection performance; and (5) conducting a comprehensive comparison of the results obtained from these models with each other and previous works. Fig. 1 shows the flowchart of our proposed work

Hence, our proposed work consists of four main stages:
1. Designing IR-BRAINNET and Modifying DEMNET [7] model
2. Preprocessing the MRI dataset and Applying SMOTE
3. Training IR-BRAINNET and Modified-DEMNET
4. Ensembling the outputs of IR-BRAINNET and Modified-DEMNET

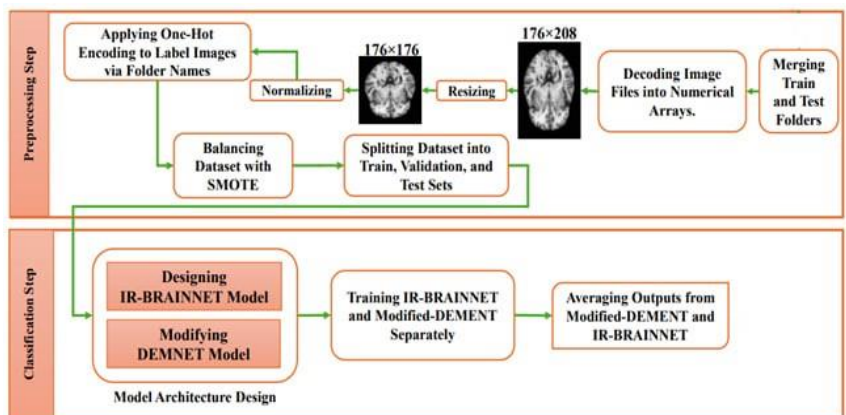

**Fig. 1** Flow chart for ensemble model

*3.2 Imaging dataset*

The dataset utilized in this study was obtained from the Kaggle platform [23], which includes four classes: MID, MOD, ND, and VMD. The images are formatted in three-channel color (RGB), with dimensions of $208 \times 176$ pixels, and saved in jpeg format. The dataset is organized into two folders, namely "train" and "test". Within the train and test folders, four directories exist, with each directory named corresponding to a class, and images belonging to each class are stored within their respective directories. A visual representation of images, each belonging to one of four classes from the Kaggle Alzheimer's dataset, is presented in Fig. 2, in Table 1, which illustrates the class distribution both before and after applying SMOTE. Without SMOTE, the dataset shows significant imbalance: 896 images for the MID class, 64 for the MOD class, 3200 for the ND class, and 2240 for the VMD class. This imbalance can hinder model performance, particularly for minority classes. After applying SMOTE, all classes were balanced with 3200 images each, addressing the disparity and enabling more effective training of the models.

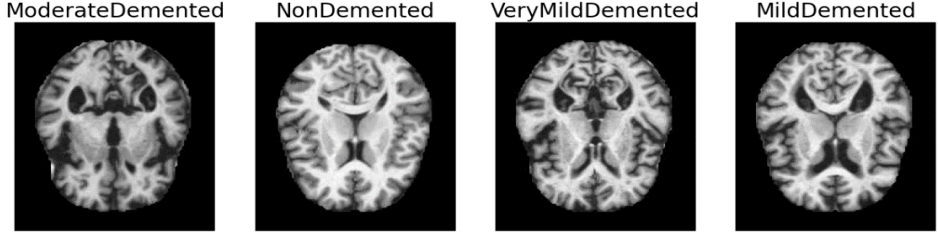

**Fig. 2** A visual representation of four images from the Kaggle Alzheimer's dataset, each belonging to one of four classes.

*3.3 Dataset preprocessing*

According to Table 1, the dataset displays a notable class imbalance, which impacts the efficacy of our proposed models, notably in precisely classifying minority classes, such as MOD, comprising only 64 images, as a result of the disproportionately higher representation of the majority class, ND, containing 3200 images. To address this issue, we used SMOTE approach to oversample the minority classes and balance the dataset. To generate new instances using SMOTE technique, we first randomly select a sample from the minority class, labeled as $x_i$. Then, we calculate its K-Nearest Neighbors and choose one of these nearest neighbors, represented by $x_{ij}$. After calculating the distance between $x_i$ and $x_{ij}$, a point is then selected on the line segment connecting these two points to generate a synthetic instance denoted as $x_{new}$. The. Eqs for creating synthetic instances using SMOTE are as follows:

$$\Delta = x_{ij} - x_i \quad (1)$$
$$x_{new} = x_i + rand[0-1] \times \Delta \quad (2)$$

In.Eq 2, $rand[0-1]$ is a random number generated between 0 and 1, determining the extent of interpolation between $x_i$ and $x_{ij}$ when creating the synthetic example $x_{new}$. Looking at. Eqs 1 and 2, incorporating interpolation between existing minority class samples and using random values from rand[0−1] generate new samples dispersed among the original minority class points. Consequently, SMOTE is a highly effective augmentation method that introduces variations into the dataset and helps reduce overfitting.

SMOTE function was provided with a vector containing all image data and another vector of labels encoded in one-hot encoding format. It was applied to the dataset with a random seed set of 42 for reproducibility, utilizing a strategy that specifically generates instances for the minority classes. Table 1 illustrates the dataset distribution after the application of SMOTE, which prioritizes generating samples for minority classes while excluding the majority class from having synthetic instances. After oversampling, the total number of images increased to 12,800. Consequently, each class now comprises 3,200 images, which aligns with the number of ND images representing the majority class.

**Table 1** Dataset distribution both before and after applying SMOTE(Without SMOTE*- With SMOTE** )

| Class | Number of Images * | Number of Images** |
|---|---|---|
| **Mild Demented (MID)** | 896 | 3200 |
| **Moderate Demented (MOD)** | 64 | 3200 |
| **Non-Demented (ND)** | 3200 | 3200 |
| **Very Mild Demented (VMD)** | 2240 | 3200 |

To facilitate the training, validation, and testing phases of the proposed CNN and evaluation of the proposed ensemble model, the dataset underwent a three-way split, allocating 70% to the training set, 10% to the validation set, and 20% to the testing set. Given the importance of downscaling images to manage spatial complexity, especially for computationally demanding techniques such as CNNs [43], all images were resized to (176×176×3) with bilinear interpolation prior to applying SMOTE and initiating CNNs training. This technique determines each new pixel value by averaging the four nearest neighbors, helping to avoid jagged edges and artifacts common with simpler methods. This is especially necessary for maintaining detail and minimizing distortions when reducing image size. Additionally, the images were normalized to a range of [0, 1] using a simplified min-max normalization method, where each pixel value is divided by 255. Normalizing images speeds up training and improves accuracy by adjusting the range and distribution of pixel values [44]. This process prevents any single pixel value from disproportionately influencing the learning process and reduces the risk of numerical instabilities.

*3.4 The First Proposed CNN Model*

The first proposed CNN in this research, IR-BRAINNET, is designed in the domain of Deep Learning for the automatic early diagnosis of AD using MRI images. The primary objective of this CNN model is to achieve high performance in detecting the early stages of AD while maintaining a simple architecture and keeping the parameter count even lower than that of DEMNET [7] to reduce computational expenses.

The structure of IR-BRAINNET consists of 6 convolutional layers, each utilizing 3×3 filters and ReLU activation function, along with 6 max-pooling layers, all with a 2×2 dimension. In detail, the convolutional layers slide 3×3 filters across the input arrays and apply ReLU activation function to generate feature maps. To reduce the number of parameters and computational costs in subsequent layers, IR-BRAINNET utilizes a downsampling method. This process involves reducing the feature maps by a factor of 2 along both width and length using max pooling layers. Convolutional and max pooling layers collaboratively extract complex features from the input images, with deeper layers capturing increasingly complex features, while initial layers focus on simpler ones. This process eliminates the need for manual feature engineering and facilitates the automatic identification of AD from MRI scans.

In IR-BRAINNET, following the last max pooling layer, the output obtained from this layer is flattened. This means that the 256 features with dimensions of 2 × 2 that emerge from the last max pooling layer are converted into a one-dimensional array with a size of 1024. They are then passed through a dense layer comprising 100 neurons with ReLU activation, which produces activation values. Subsequently, these activation values are fed into another dense layer with 4 neurons, where the output is transformed using the Softmax activation function from raw scores into a probability distribution. This probability distribution not only facilitates training by adjusting network parameters to minimize differences between predicted and actual class labels but also serves as a basis for evaluating the model. This evaluation occurs both at the end of each training round in the validation phase and during the testing phase. The formulations for the ReLU and Softmax functions are represented by. Eq 3 and.Eq 4, respectively:

$$\text{ReLU}(X) = \text{Max}(0, X) \qquad (3)$$

$$\text{Softmax} = \left(\frac{e^{S_N}}{\sum_j^c e^{S_j}}\right) \qquad (4)$$

In.Eq 3 for any input value $X$, if $X$ is greater than 0, the output is simply $X$. However, if $X$ is less than or equal to 0, the output becomes 0. ReLU is considered a non-linear function due to its output's non-linear relationship with its input. Specifically, when the input $X$ is positive, the output maintains a linear relationship with the input, as it equals the input value. Conversely, when $X$ is non-positive, the output becomes 0, breaking this linearity. This property allows ReLU to introduce non-linearity into neural networks, aiding in learning complex patterns within the data.

In.Eq 4, $S_N$ denotes the unprocessed score or logit for the Nth class, while $e^{S_N}$ represents the exponential of this unprocessed score corresponding to the Nth class. The term $\sum_j^c e^{S_j}$ calculates the sum of exponentials of unprocessed scores for all classes (j) in the output layer, where c signifies the total number of classes.

The total number of parameters in IR-BRAINNET, which includes two dense layers in its fully connected layer alongside the six convolutional layers, amounts to 1,801,464. To leverage insights from the pre-trained VGG-19 model, which was trained on the ImageNet dataset [26], we initialized the weights of the second convolutional layer in our IR-BRAINNET by transferring the weights from the third convolutional layer of VGG-19, containing 73,856 parameters, to the second layer of the IR-BRAINNET model with the same 73,856 parameters. We maintained all 1,801,464 parameters of IR-BRAINNET as trainable. This allowed us to fine-tune the parameters of the second layer of IR-BRAINNET, which utilized pre-trained weights as its weight initializers.

Transfer learning can address challenges stemming from insufficient labeled training data and can be employed as a strategy to mitigate overfitting [45]. It is possible to enhance performance by transferring pre-trained knowledge, even though ImageNet [26] contains different classes from those used in AD detection. Networks' early layers capture basic features like edges, while deeper layers build on these to develop more intricate, task-specific representations. Since these fundamental features are typically consistent across different image categories, utilizing this knowledge can improve performance and generalization in AD detection.

In comparison to DEMNET [7], which aims for a similar low CNN parameter countIn comparison to DEMNET[7], which aims for a similar low CNN parameter count, IR-BRAINNET exhibits a significantly smaller parameter count in terms of both total parameters and trainable parameters. DEMNET totals 4,534,996 parameters, of which 4,532,628 are trainable.

*3.5 The Second Proposed CNN Model*

In line with our research objectives, we developed a second model, Modified-DEMNET, which leverages the convolutional block structure of the original DEMNET [7] model. We modified the third layer of the DEMNET model and the last dimension reducer in the final block of the DEMNET network by replacing the max pooling 2D with the average pooling 2D. The key distinction between max pooling 2D and average pooling 2D with a 2×2 window is in how they handle values within the specified region. In max pooling 2D, the highest value is selected from the four numbers in the 2×2 window, while in average pooling 2D, the average value of these four numbers is chosen. In medical image analysis, lesions or abnormalities often show significant variability in their spatial distribution across images [46]. For instance, a lesion might appear in only a small part of the image, where max pooling is particularly effective. On the other hand, if the lesion is spread across a larger portion of the image, average pooling can be more beneficial. By incorporating both max pooling and average pooling in Modified-DEMNET, the model leverages each technique's strengths to effectively capture localized and global features. This approach enhances the model's ability to generalize across diverse image data, reducing the likelihood of learning patterns that are specific to the training set and mitigating the risk of overfitting.

Moreover, we simplified the fully connected layer of DEMNET into two layers: a parameter-free flattened layer and a Dense layer with 100 neurons, using ReLU activation function to extract the 100 activation values. Similar to IR-BRAINNET, and considering the four classes in the Kaggle Alzheimer's dataset, a Dense layer with four neurons that uses the Softmax activation function was employed as the prediction layer. In Modified-DEMNET, we kept all the batch normalization layers intact, as they help prevent overfitting and address issues like vanishing and exploding gradients. By stabilizing the input distributions to each layer—distributions that might otherwise fluctuate during training—batch normalization ensures more uniform activations across the network. This consistency helps address the issues of vanishing and exploding gradients by keeping the activation ranges stable. Moreover, batch normalization introduces a level of noise by basing the normalization of inputs on the mean and variance from a mini-batch instead of the entire dataset. This randomness, similar to the effect of dropout, reduces overfitting by preventing the model from relying too heavily on specific patterns in the training data. Through the adaptation of the DEMNET architecture and implementing key modifications tailored to our objectives, we aimed to attain promising results in AD detection while reducing the total and trainable parameters of the DEMNET model.

*3.6 Weight Initialization Strategy for IR-BRAINNET and Modified-Dement*

In neural networks, weights are essential because they significantly affect the network's performance; therefore, considerable effort must be invested in determining their initial values and configurations [47]. The choice of initial weights in neural networks is crucial as it can help prevent issues such as gradient explosion or vanishing gradients. [48] In this paper, we used the Kaiming He weight initialization method to properly initialize the weights of all trainable layers in both CNNs, except for the second layer of IR-BRAINNET. According to [48], the formula for the Kaiming He initializer is as follows:

$$W \sim \mathcal{N}\left(0, \frac{2}{n_{in}}\right) \quad (5)$$

Here, $\mathcal{N}$ represents the normal distribution, $n_{in}$ denotes the number of input units, and $W$ is the weight matrix of a layer. According to.Eq 5, The Kaiming He initialization method sets the weights of a layer by drawing them from a normal distribution with a mean of 0 and a variance of $\frac{2}{n_{in}}$. As a result, this method ensures that the variance of activations stays uniform throughout the network layers. By appropriately scaling the weight variances, this initialization method effectively guards against gradients becoming excessively small (vanishing) or excessively large (exploding).

We incorporated ReLU activation in both CNNs to introduce non-linearity into the models. The main reason for choosing Kaiming He initialization in this study is that unlike Xavier initialization (also known as Glorot uniform), which assumes linear activations and may not perform well with ReLU, Kaiming He initialization is specifically designed to address the non-linearity of ReLU functions [49]. According to [49], Kaiming He initialization enables the training of very deep rectified models from scratch and allows for exploring more complex network architectures, whether deeper or wider. Thus, this method is beneficial for two low-parameter networks with ReLU activations. For example, one issue that can arise with ReLU activations if weights are not initialized correctly is the problem of dead neurons. This occurs when a neuron fails to activate and consistently outputs zero for any negative input. Proper initialization with Kaiming helps avoid this problem.

Therefore, Kaiming He initialization ensures stable gradient propagation throughout both CNNs, avoiding the problem of dead neurons, which allows neurons to effectively contribute to feature representation and backpropagation. This enhances learning, improves generalization, and potentially reduces overfitting.

## 3.7 The Proposed Ensemble Model

To further enhance the identification of AD stages from MRI images, we combined predictions from IR-BRAINNET and Modified-DEMNET to form the ensemble model's prediction. As previously mentioned, our ensembling approach involves taking the outputs of CNNs, obtained using Softmax activation function, and feeding them into the average layer. The. Eq for the average function, which combines the outputs of the individual CNNs, is as follows:

$$\text{Average output} = \frac{1}{n} \sum_{i=1}^{n} Output_{CNN_i} \quad (6$$

In this.Eq, $Output_{CNN_i}$ refers to the output of the $i^{th}$ CNN. The sum is then taken over all n CNNs (in this study, n is two), and the result is divided by the number of n. Given the distinct architectural characteristics of IR-BRAINNET and Modified-DEMNET, making use of the diverse perspectives offered by each model and consolidating their outputs into a single prediction through averaging fosters a complementary relationship between the models, which can enhance AD detection capabilities. This integration capitalizes on the strengths of both models and can effectively produce a more robust and accurate prediction framework for AD detection. Additionally, our proposed ensemble model effectively reduces noise variance, leading to more stable and reliable predictions with fewer fluctuations. As a result of this variance reduction, the model generalizes better to new, unseen data and is less prone to overfitting.

**Algorithm 1** The pseudocode of our proposed ensemble model

| **Algorithm 1: Psuedo-Algorithm of proposed ensemble model** |
|---|
| **Input:** |
|     Images: MRI images |
|     Labels: Labels |
|     IR-BRAINNET: our first proposed CNN model |
|     Modified-DEMNET: our second proposed CNN model |
| **Output:** |
|     EnsemblePrediction: An aggregated prediction using an Average Layer. |
| 1    ImagesSize ← (176×176×3) |
| 2    Prepcoess ← (resize(ImagesSize)) and normalize) |
| 3    $X_{SMOTE}, Y_{SMOTE}$ ← SMOTE(Images, Labels) |
| 4    X, Y ← Preprcoess($X_{SMOTE}, Y_{SMOTE}$) |
| 5    $X_{Train}, Y_{Train}, X_{Val}, Y_{Val}, X_{Test}, Y_{Test}$ ← split(X,Y) |
| 6    Train(IR-BRAINNET,epochs=50,optimizer=Adam,learningRate=0.0001) using $X_{Train}, Y_{Train}, X_{Val}, Y_{Val}$ |
| 7    Train(Modified-DEMNET,epochs=50,optimizer= Adam,learningRate=0.0001) using $X_{Train}, Y_{Train}, X_{Val}, Y_{Val}$ |
| 8    EnsembleSet ← {IR-BRAINNET, Modified-DEMNET} |
| 9    AverageLayerInputs ← Compute the output of the model using $X_{Test}$ for model in EnsembleSet |
| 10    EnsemblePrediction ← Compute the average of all elements in AverageLayerInputs |

The operation of averaging the probability vectors derived from the Softmax function of IR-BRAINNET and Modified-DEMNET to form the ensemble probability vector is illustrated in Fig 3. In Fig 3, the probability vectors $P_{MID}^{IR-BRAINNET}, P_{MOD}^{IR-BRAINNET}, P_{ND}^{IR-BRAINNET}, P_{VMD}^{IR-BRAINET}$ denote the likelihoods assigned by IR-BRAINET for categorizing an instance into the classes MID, MOD, ND, and VMD, respectively. Similarly, vectors $P_{MID}^{Modified-DEMNET}, P_{MOD}^{Modified-DEMNET}, P_{ND}^{Modified-DEMNET}$, and $P_{VMD}^{Modified-DEMNET}$ denote the probabilities for Modified-DEMENT. The ensemble model's probability vectors are denoted as $P_{MID}^{Ensemble}, P_{MOD}^{Ensemble}, P_{ND}^{Ensemble}, P_{VMD}^{Ensemble}$, which reflects the aggregated likelihood of assigning an instance to each class.

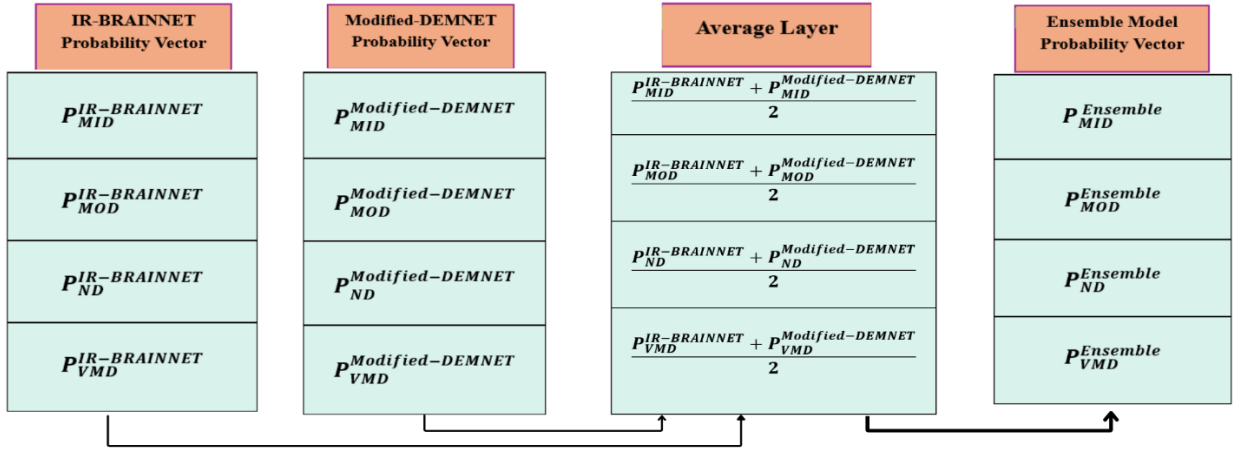

**Fig. 3** Illustration of the averaging process of two probability vectors generated by the proposed CNNs

## 4. Experimental Results

In the Evaluation Results, we analyzed the performance of IR-BRAINNET, Modified-DEMNET, and our proposed ensemble model in detecting the early stages of AD. This section outlines the Performance Setup, Performance Metrics. Eqs, and training phase configurations, and presents the outcomes based on the likelihood vectors of models generated during the testing phase.

*4.1 Performance Setup*

The study was carried out using the Python programming language within the Google Colab environment [50]. For designing and training our deep learning models, we harnessed the Tensorflow framework alongside its Keras package. The computational resources afforded by the complimentary cloud service in Google Colab encompassed an Nvidia Tesla T4 GPU boasting 15.6 GB memory, 12.6 GB RAM, an Intel Xeon CPU operating at 2.20 GHz, and 107.7 GB HDD

*4.2 Model Design and Training*

After the preprocessing step, we proceeded to train the proposed CNNs by inputting the preprocessed image arrays into them. Both models underwent 50 epochs of training, during which their parameters were iteratively adjusted using the adaptive moment estimation (Adam) optimizer. The Adam optimizer utilizes the models' loss to update their parameters based on the gradients of the loss function. For the multi-class classification task, we employed the Categorical Cross-Entropy loss function to measure the dissimilarity between the true distribution of the data and the predicted distribution. We initiated training with an initial learning rate of 0.0001 for both CNNs and, as a strategy for ensuring stable convergence and reducing the risk of overfitting, continuously monitored their loss values throughout the optimization process. If there was no discernible decrease in loss after 3 epochs, we automatically proceeded to reduce the learning rate by multiplying it by 0.1. We specified the exponential decay rate for the first moment estimate as 0.9 and the exponential decay rate for the second moment estimate as 0.999. Additionally, to prevent division by zero and ensure stability during training, we set the small constant epsilon to 1e-07. After training both models, we designed our proposed ensemble model by merging the last prediction layers of these models. The merging process involves utilizing the average layer, resulting in our ensemble model incorporating the trained parameters of both CNNs.

*4.3 Evaluation Metrics*

In our model evaluation process, we relied on metrics such as accuracy, precision, recall, and F1-score. To provide clarity on these metrics, we must first define True Positives (TP), False Positives (FP), True Negatives (TN), and False Negatives (FN). For clarification, let's consider the prediction vector of a specific model, like IR-BRAINNET, where, for each image, a class exists that IR-BRAINNET predicted the image belongs to. When evaluating the model for the MOD class, TP represents instances correctly identified as MOD, FP involves cases incorrectly labeled as MOD, TN indicates instances correctly recognized as not belonging to MOD, and FN consists of cases inaccurately classified as not being part of the MOD class.

The accuracy metric evaluates the correctness of our models in predicting both true positives and negatives. Its calculation is determined by the. Eq:

$$\text{Accuracy} = \frac{TP + TN}{TP + FP + TN + FN} \tag{7}$$

Precision is another crucial metric that gauges the ratio of true positive observations to total positive predictions. The precision can be calculated using the. Eq:

$$\text{Precision} = \frac{TP}{TP + FP} \tag{8}$$

Recall, also known as sensitivity, quantifies the classifier's capability to identify all positive samples, and it is calculated using the following. Eq:

$$\text{Recall} = \frac{TP}{TP + FN} \tag{9}$$

The F1-score is the harmonic mean of precision and recall, representing how well the model balances between precision and recall. It is calculated using the. Eq:

$$F1 - \text{score} = \frac{2 \times \text{Precision} \times \text{Recall}}{\text{Precision} + \text{Recall}} \tag{10}$$

## 4.4 Evaluation

In this section, we delve into the experimental results of IR-BRAINNET, Modified-DEMNET, and our proposed ensemble model for the early stages of AD classification. Upon analyzing the test prediction vectors alongside the test labels encoded in one-hot format, and based on the TP, FP, TN, and FN, we calculated the test accuracy, precision, recall, and F1-Score for each model in both the NO-SMOTE and SMOTE scenarios. The performance metrics of the models under both NO-SMOTE and SMOTE scenarios are presented in Table 2.

According to Table 2, in the absence of SMOTE and synthetic instances, the proposed ensemble model displayed superior performance compared to its individual CNNs. It achieved the highest test accuracy, precision, recall, and F1-Score at 98.28%, 98.80%, 96.28%, and 97.49% respectively. In contrast, IR-BRAINNET achieved lower results than the ensemble model, with an accuracy of 97.27%, precision of 97.86%, recall of 95.44%, and F1-Score of 96.60%. Furthermore, Modified-DEMNET exhibited metrics that were inferior to both IR-BRAINNET and the ensemble model, with an accuracy of 95.55%, precision of 96.95%, recall of 96.23%, and F1-Score of 96.57%.

Transitioning to the SMOTE scenario, notable advancements were observed across all models. IR-BRAINNET achieved significantly higher metrics, with test accuracy at 99.80%, precision at 99.80%, recall at 99.80%, and F1-Score at 99.80%. Modified-DEMNET also demonstrated improvements, reaching accuracy metrics at 99.73%, precision at 99.72%, recall at 99.72%, and F1-Score at 99.72%. The ensemble model performed exceptionally well in the SMOTE scenario, achieving the highest metrics: test accuracy at 99.92%, precision at 99.92%, recall at 99.92%, and F1-Score at 99.92%. With these results, it surpasses its performance in the NO-SMOTE scenario.

In both scenarios, the ensemble model consistently showcased superior efficacy over its individual CNN components. This demonstrates its effectiveness in bolstering overall detection performance by consolidating a unified probability distribution vector through the averaging of CNN outputs. Additionally, the utilization of SMOTE further augmented accuracy and predictive capabilities across all three models by effectively balancing the dataset through the generation of synthetic instances. It is also noteworthy that IR-BRAINNET demonstrates superior performance based on accuracy, precision, recall, and F1-Score compared to Modified-DEMNET, positioning it as the second-best performing model in both scenarios, despite having a lower number of parameters compared to Modified-DEMNET.

**Table 2** *Evaluation results for IR-BRAINNET, Modified-DEMNET, and proposed ensemble model*

| Model | Accuracy (NO-SMOTE) | Precision (NO-SMOTE) | Recall (NO-SMOTE) | F1-SCORE (NO-SMOTE) | Accuracy (SMOTE) | Precision (SMOTE) | Recall (SMOTE) | F1-SCORE (SMOTE) |
|---|---|---|---|---|---|---|---|---|
| **IR-BRAINNET** | 0.97 | 0.97 | 0.95 | 0.96 | 0.99 | 0.99 | 0.99 | 0.99 |
| **Modified-DEMNET** | 0.95 | 0.96 | 0.96 | 0.96 | 0.99 | 0.99 | 0.99 | 0.99 |
| **Ensemble Model** | 0.98 | 0.98 | 0.96 | 0.97 | 0.99 | 0.99 | 0.99 | 0.99 |

We have presented the learning curves of IR-BRAINNET and Modified-DEMNET during training on both balanced and imbalanced datasets. Fig. 4 illustrates the learning curves during training on both the balanced dataset (SMOTE scenario) and the imbalanced dataset (NO-SMOTE scenario). In both cases, the validation accuracy and loss curves consistently align with the trends observed in the training accuracy and loss curves, confirming the absence of overfitting.

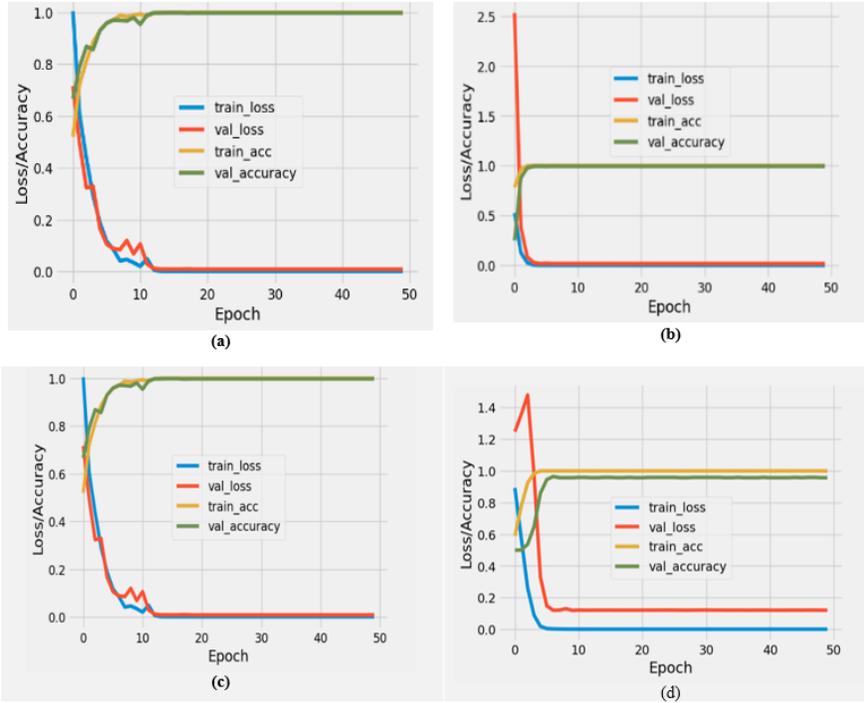

**Fig. 4** Visual results the learning curves **(a)**IR-BRAINNET learning curve (SMOTE), **(b)**Modified-DEMNET learning curve(SMOTE), **(c)**IR-BRAINNET learning curve(NO-SMOTE) **(d)**Modified-DEMNET learning curve(NO-SMOTE)

We also provided the confusion matrix of our proposed ensemble model in both scenarios, derived from the predictions made by the ensemble model when the test dataset is inputted into it. Fig. 5 illustrates the confusion matrix for our ensemble model in the SMOTE scenario. Upon examining Fig. 5, it becomes evident that our proposed ensemble model accurately classified all instances belonging to the minority class (MOD) in the SMOTE scenario and nearly all instances (with only one false negative) in the scenario without applying SMOTE. Furthermore, owing to the balanced dataset in the SMOTE scenario, our proposed ensemble model demonstrates superior performance, with only two instances incorrectly classified.

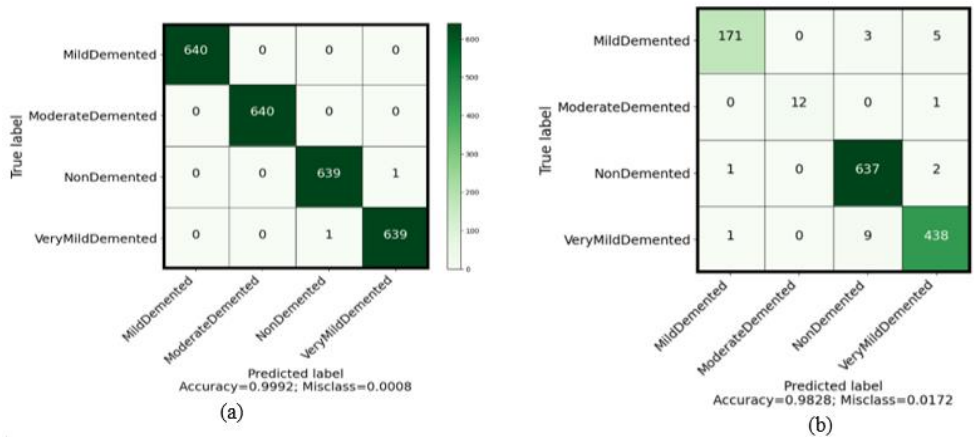

**Fig. 5** Visual results the confusion matrix of **(a)** Confusion Matrix for the proposed Ensemble Model (SMOTE), **(b)**Confusion Matrix for proposed Ensemble Model (NO-SMOTE)

We additionally plotted the ROC curves for all models. Fig. 8 illustrates the ROC curves for IR-BRAINNET, Modified-DEMNET, and our proposed ensemble model in both the SMOTE and NO-SMOTE scenarios. As evident in Fig. 6, our proposed ensemble model outperforms the other two models in both scenarios, with an AUC of 99.93% for the SMOTE scenario and an AUC of 97.70% for the NO-SMOTE scenario. IR-BRAINNET has an AUC of 99.89% and 97.07% in the SMOTE and NO-SMOTE scenarios, respectively, while Modified-DEMNET achieves an AUC of 99.82% for the SMOTE scenario and 97.52% in the NO-SMOTE scenario. Moreover, Fig. 6 shows that the ROC curves of the ensemble model are closer to the top-left corner in both scenarios, indicating higher performance compared to its individual CNNs.

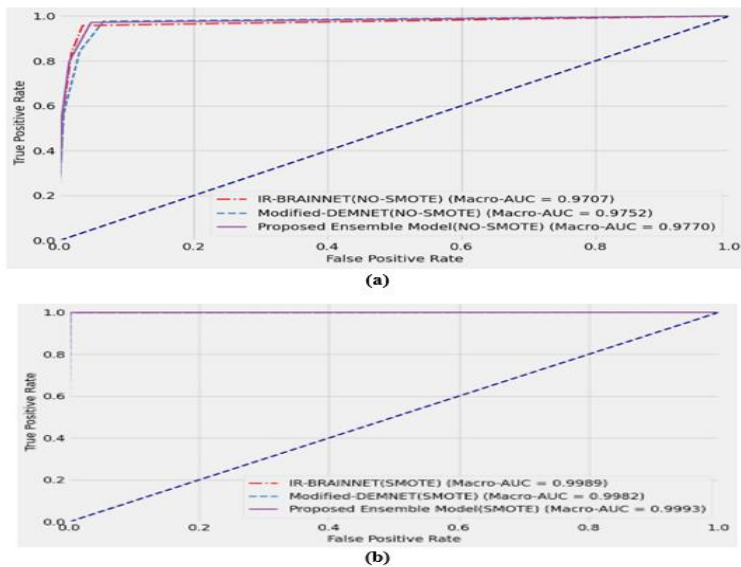

**Fig. 6** Visual results of the ROC curves **(a)**ROC curves for all models (SMOTE), **(b)**ROC curves for all models (NO-SMOTE)

**Table 3** The evaluation results for the proposed ensemble SMOTE and NO-SMOTE scenarios, model across different classes

| Class | Precision (SMOTE) | Recall (SMOTE) | F1-Score (SMOTE) | Precision (NO-SMOTE) | Recall (NO-SMOTE) | F1-Score (NO-SMOTE) |
|---|---|---|---|---|---|---|
| MildDemented | 1.00 | 1.00 | 1.00 | 0.98 | 0.95 | 0.97 |
| ModerateDemented | 1.00 | 1.00 | 1.00 | 1.00 | 0.92 | 0.96 |
| NonDemented | 0.99 | 0.99 | 0.99 | 0.98 | 0.99 | 0.98 |
| VeryMildDemented | 0.99 | 0.99 | 0.99 | 0.98 | 0.97 | 0.97 |

In Table 3, we presented the evaluation results for our proposed ensemble model, individually for each of the four classes, in both the SMOTE and NO-SMOTE scenarios.

As evident from Table 3, the precision of the minority class (MOD) is at 100% in the SMOTE scenario and nearly 100% in the NO-SMOTE scenario. A precision of 100% indicates that every prediction made by the model for the MOD class is accurate, with no false positives within this category. Also, based on recall, in the NO-SMOTE scenario, our proposed ensemble model exhibits a recall of 92.30%, while for SMOTE, it is at 100%. The high recall rates for the MOD class indicate that the model is highly effective in not missing positive instances of this minority class, even in an imbalanced situation. Missing instances of the MOD class, which could lead to early treatment of AD disease, highlights the crucial significance of the model's performance in this regard. Regarding other classes in the NO-SMOTE scenario, the model achieved a precision of 98.84%, a recall of 95.53%, and an F1-Score of 97.16% for MildDemented individuals. Similarly, in the NonDemented category, the model demonstrated a precision of 98.15%, a recall of 99.53%, and an F1-Score of 98.84%. Lastly, for the VeryMildDemented class, the model yielded a precision of 98.21%, a recall of 97.77%, and an F1-Score of 97.99%. Furthermore, in the SMOTE scenario, our proposed ensemble model demonstrates outstanding performance, achieving 100% accuracy, precision, recall, and F1-score not only for the minority class MOD but also for MID, which represents the second minority class after MOD. The ensemble model's proficiency in achieving flawless metrics for both minority classes under the influence of SMOTE underscores its ability to learn representations of these underrepresented classes. With the help of SMOTE, which provided the proposed ensemble model (and its CNNs) with more instances to learn, the model gained an effective understanding of these classes. This indicates the ability of the ensemble model to extract effective features and highlights the effectiveness of SMOTE in the way it generates instances. Regarding NonDemented and VeryMildDemented in the SMOTE scenario, the model achieved a precision, recall, and F1-Score of 99.84% for the NonDemented class. Similarly, for the VeryMildDemented category, the model demonstrated precision, recall, and F1-Score values also at 99.84%.

The consistent and high performance across both balanced and imbalanced distributions, even for MOD, demonstrates the reliability and effectiveness of the ensemble model in providing accurate diagnoses. This capability can lead to effective early AD detection and optimal treatment.

We calculated the Floating Point Operations (FLOPs) to assess the computational complexity and efficiency of the proposed models. FLOPs measure the computational demands and speed of model operations. For convolutional layers, FLOPs are calculated by multiplying the input feature map size by the input/output channels, kernel size, and 2. Pooling layers involve multiplying the input feature map size by the input channels and kernel size, while fully connected layers involve multiplying the input and output feature dimensions by 2. Summing the FLOPs for all layers provides the total FLOPs for a model.

Our analysis shows that Modified-DEMNET has the lowest computational complexity at 0.5155 GFLOPs, demonstrating superior efficiency. IR-BRAINNET requires 2.8071 GFLOPs, while the proposed ensemble model demands the most at 3.3226 GFLOPs. This highlights the efficiency of our models compared to more complex CNNs like VGG-16 and VGG-19, which require 9.5536 GFLOPs and 12.1232 GFLOPs, respectively, for an input shape of (176,176,3). Additionally, the original DEMNET model has 0.5182 GFLOPs, exceeding Modified-DEMNET by 2,710,686 FLOPs.

One key reason why IR-BRAINNET has a higher GFLOPs count than Modified-DEMNET is the sharp increase in the number of filters in its early layers. In IR-BRAINNET, the filters jump from 64 in the first layer to 128 in the second layer, significantly contributing to its computational complexity. In contrast, Modified-DEMNET features a gradual increase in filters, progressing from 16 to 256 across layers. Calculating the GFLOPs for the second convolutional layer of IR-BRAINNET (with 128 filters) reveals a value of 2.2858 GFLOPs, highlighting this layer's substantial impact on its complexity.

Despite the computational challenges, this design enhances IR-BRAINNET's performance. The sharp increase in filters enables the network to capture complex feature patterns in deeper layers, creating a richer representation of input data. This allows IR-BRAINNET to extract detailed and nuanced features, enabling its deeper layers to learn intricate and abstract patterns. In contrast, Modified-DEMNET, with fewer filters early on, generates less detailed features, potentially limiting the complexity of patterns its deeper layers can learn.

Consequently, IR-BRAINNET achieves improved performance, with an accuracy of 97.26% compared to Modified-DEMNET's 95.54% in the NO-SMOTE scenario. This strategic use of pre-trained VGG-19 knowledge in its early layers further enhances IR-BRAINNET's feature representation abilities, leveraging its complexity for superior performance.

The increased complexity of our proposed ensemble model is a characteristic shared by all ensemble methods. By averaging outputs, adding more CNNs can reduce variance and enhance performance, albeit with an increase in FLOPs due to the higher computational load. However, our ensemble model offers a computational advantage through an efficient algorithm that averages output vectors from two CNNs using only addition and division by 2, requiring negligible FLOPs for this operation.

Table 4 compares training and prediction times for all models, showing that all exhibit efficient prediction times. Training times for IR-BRAINNET and Modified-DEMNET are efficient, even with dataset variations. In the NO-SMOTE scenario (4,608 images), IR-BRAINNET's training time per epoch is 22 seconds, while Modified-DEMNET's is 8 seconds. For the SMOTE scenario (9,216 images), IR-BRAINNET requires 40 seconds per epoch, and Modified-DEMNET needs 17 seconds, reflecting increased training time due to a larger dataset. Training time per batch (size 32) is 138 milliseconds for IR-BRAINNET and 58 milliseconds for Modified-DEMNET, while prediction time is 32 milliseconds and 11 milliseconds, respectively. The ensemble model's prediction time is 56 milliseconds, slightly higher than the combined time of IR-BRAINNET and Modified-DEMNET but still efficient.

The time complexity for averaging the outputs of two CNNs with four neurons in the prediction layer and T images is O(2 × 4 × T), simplifying to O(T). This linear growth ensures the ensemble method remains computationally efficient despite its added complexity.

Table 4 Comparison of Training and Prediction Times Across Models

| Model | Training time (NO-SMOTE) | Training time (SMOTE) | Training/batch (32) | Prediction/batch (32) |
|---|---|---|---|---|
| IR-BRAINNET | 22s | 40s | 138 ms | 32ms |
| Modified-DEMNET | 8s | 17s | 58 ms | 11ms |
| Proposed Ensemble model | - | - | - | 56ms |

We estimated memory consumption based on the sum of trainable and non-trainable parameters. For IR-BRAINNET, the total memory consumption is approximately 6.87 MB, while Modified-DEMNET consumes about 6.95 MB. Both models exhibit low memory consumption, with IR-BRAINNET using slightly less memory. The proposed ensemble model, which combines both IR-BRAINNET and Modified-DEMNET, has a total memory consumption of 13.82 MB. The original DEMNET model [7] has a memory consumption of 17.29 MB, which is even greater than that of our proposed ensemble model.

## 5. Comparison with Other Models

We have also conducted a comparison of our proposed models with other previous models from the literature. Detailed results of this comparison are available in Table 5. Table 5 clearly illustrates that IR-BRAINNET, Modified-DEMNET, and our proposed ensemble model demonstrate superior performance in terms of accuracy, precision, recall, and F1-score on the Kaggle Alzheimer's dataset in both SMOTE and NO-SMOTE scenarios compared to previous models by other authors who used this dataset.

Table 5 Comparison with other models

| Model | Type of Classifications | Accuracy | Precision | Recall | F1-score |
|---|---|---|---|---|---|
| **IR-BRAINNET**** | **Multi** | **0.99** | **0.99** | **0.99** | **0.99** |
| **Modified-DEMNET**** | **Multi** | **0.99** | **0.99** | **0.99** | **0.99** |
| **Proposed Ensemble Model* *** | **Multi** | **0.99** | **0.99** | **0.99** | **0.99** |
| **IR-BRAINNET*** | **Multi** | **0.97** | **0.97** | **0.95** | **0.96** |
| **Modified-DEMNET*** | **Multi** | **0.95** | **0.96** | **0.96** | **0.96** |
| **Proposed Ensemble Model*** | **Multi** | **0.98** | **0.98** | **0.96** | **0.97** |
| ADD-Net ** [31] | Multi | 0.97 | 97 | 97 | 0.97 |
| ADD-Net *[31] | Multi | 0.66 | 82 | 89 | 0.84 |
| DEMENT**[7] | Multi | 0.95 | 96 | 95 | 95 |
| DEMENT*[7] | Multi | 85 | 80 | 88 | 83 |
| CNN [15] | Binary | 0.99 | 0.99 | 0.99 | 0.99 |
| CNN-LSTM [39] | Binary | 0.99 | 100.00 | 0.99 | 0.99 |
| TLBO based DeepCNN [37] | Multi | 0.95 | - | 0.96 | 0.96 |

## 6. Conclusion and Future Scopes

In this study, we proposed three models based on convolutional layers for the early detection of AD: IR-BRAINNET, Modified-DEMNET, and an ensemble model. The ensemble model combines the outputs of IR-BRAINNET and Modified-DEMNET using the average function during the aggregation process. We utilized the Kaggle Alzheimer's dataset for both training and evaluation phases. This dataset shows a notable imbalance, and to address potential issues stemming from this disparity, we utilized SMOTE to generate synthetic instances. Both CNNs underwent 50 epochs of training, and in the evaluation phase, all three proposed models showcased remarkable performance across key metrics, including accuracy, recall, precision, and F1-score, even when dealing with the imbalanced dataset. The ensemble model consistently outperformed its constituent CNNs in both scenarios. Notably, it achieved 100% precision for the minority class MOD in the SMOTE scenario and maintained nearly 100% precision for this class, with just one false positive, in the NO-SMOTE scenario. Moreover, in the SMOTE scenario, the ensemble model displayed remarkable classification performance, correctly identifying all instances without any false negatives or false positives for both MOD and MID. Regarding CNNs, the proposed IR-BRAINNET showcased the second-highest performance and superior AD detection proficiency compared to Modified-DEMENT in both scenarios.

In our upcoming research, we aim to explore the utilization of a diverse range of datasets, including ADNI, OASIS, and other relevant datasets, for training both IR-BRAINNET and Modified-DEMNET. Subsequently, we will test these CNNs and the proposed ensemble model on these datasets. This approach will enable us to conduct a comprehensive evaluation of the performance and adaptability of both models across a wide range of data sources. Moreover, leveraging the pre-trained weights of CNNs trained on the Kaggle Alzheimer's dataset as the initializers for CNN weights and subsequently fine-tuning all layers or part of the layers of CNNs on ADNI, OASIS, and other datasets could enhance the performance of models for deployment in broader contexts.

Additionally, we aim to extend the training process to incorporate our proposed ensemble model as a separate entity, rather than solely as a model for combining votes. By treating the ensemble model as an independent entity and training it alongside the individual models, we intend to further improve its performance. Focusing on the effectiveness of the ensemble framework and ensemble logic during the backpropagation process will be a key aspect of this effort.

This study utilized a linear ensemble operator (average function), which does not account for the interactions among our CNNs. To fully use the diverse architectural components within our ensemble model in future work, we are considering incorporating non-linear functions, such as the Sugeno and Choquet fuzzy integral functions, to capture better and leverage interactions among CNNs. An additional approach for introducing more diversity in our future work is to incorporate pre-trained deep CNN architectures such as VGG-Net, ResNet, and Dense-Net into our ensemble model. This, coupled with the incorporation of the fuzzy Choquet integral and Sugeno integral, facilitates the dynamic management and harnessing of architectural diversity, which ultimately can result in the effective identification of various stages of AD.

**Funding** The authors assert that they did not receive any funding, grants, or additional support while preparing this manuscript.

**Data availability** *https://www.kaggle.com/datasets/tourist55/alzheimers-dataset-4-class-of-images*

**Declarations**

**Conflict of interest** The authors declare that they have no conflict of interest.